\documentclass[sigconf]{acmart}
\usepackage[utf8]{inputenc}

\usepackage{graphicx,subfigure,icomma}
\usepackage{listings}
\usepackage[simplified]{pgf-umlcd}
\usepackage{svg}
\svgsetup{inkscapelatex=false}

\usepackage{multirow}
\usepackage{amsmath}
\usepackage{xcolor}

\usepackage{algorithm}
\usepackage[noend]{algpseudocode}

\AtBeginDocument{%
  }

\copyrightyear{2022}
\acmYear{2022}
\setcopyright{acmcopyright}\acmConference[MODELS '22 Companion]{ACM/IEEE 25th International Conference on Model Driven Engineering Languages and Systems}{October 23--28, 2022}{Montreal, QC, Canada}
\acmBooktitle{ACM/IEEE 25th International Conference on Model Driven Engineering Languages and Systems (MODELS '22 Companion), October 23--28, 2022, Montreal, QC, Canada}
\acmPrice{15.00}
\acmDOI{10.1145/3550356.3561592}
\acmISBN{978-1-4503-9467-3/22/10}

\title{Towards Automatically Extracting UML Class Diagrams from Natural Language Specifications}

\author{Song Yang}
\affiliation{%
  \institution{Université de Montréal}
  \city{Montreal}
  \country{Canada}}
\email{song.yang.1@umontreal.ca}
\author{Houari Sahraoui}
\affiliation{%
  \institution{Université de Montréal}
  \city{Montreal}
  \country{Canada}}
\email{houari.sahraoui@umontreal.ca}
\date{September 2022}
\renewcommand{\shortauthors}{Yang et al.}

\begin{abstract}
In model-driven engineering (MDE), UML class diagrams serve as a way to plan and communicate between developers. However, it is complex and resource-consuming. We propose an automated approach for the extraction of UML class diagrams from natural language software specifications. To develop our approach, we create a dataset of UML class diagrams and their English specifications with the help of volunteers. Our approach is a pipeline of steps consisting of the segmentation of the input into sentences, the classification of the sentences, the generation of UML class diagram fragments from sentences, and the composition of these fragments into one UML class diagram. We develop a quantitative testing framework specific to UML class diagram extraction. Our approach yields low precision and recall but serves as a benchmark for future research.
\end{abstract}

\begin{document}

\maketitle

\section{Introduction}
Software development is a complex and error-prone process. Part of this complexity comes from the gap between domain experts who are familiar with the domain knowledge but have limited expertise with development tools, and software specialists who master the development environments but are unfamiliar with the target application domain. To fill that gap, the model-driven engineering paradigm aims at raising the level of abstraction in development activities by considering domain models, such as UML class diagrams, as first-class development artifacts.

Though smaller, a gap still exists between the domain concepts and the tools and languages that are produced to model them~\cite{mussbacher2020opportunities}. For a domain specialist, creating UML models from scratch is a time-consuming and error-prone process that requires various technical skills. To address that problem, various approaches target the generation of models from different structured information such as user stories~\cite{elallaoui2015automatic}. However, little work has been done on the extraction of natural language specifications. In the specific case of UML class diagrams, existing work rely either on techniques that use machine learning in a semi-automated process~\cite{Saini:2022} or rule-based techniques that are fully automated but require a restricted input~\cite{Abdelnabi:2020generating, More:2012}. In this paper, we propose an approach that combines both machine learning and rules while accepting free-flowing text.

Our approach uses natural language patterns and machine learning to fully automate the generation process. We first decompose a specification into sentences. Then, using a trained classifier, we tag each sentence as describing either a class or a relationship. Next, using grammar patterns, we map each sentence into a UML fragment. Finally, we assemble the fragments into a complete UML diagram using a composition algorithm. In addition to our approach, we build a dataset thanks to the effort of the modeling community. This dataset is used to train the classifier et to evaluate the approach.

We evaluate our approach on a dataset of 62 diagrams containing 624 fragments. Although the accuracy of our approach does not reach an accuracy level needed for practical use, our work explores the benefits of mixing machine learning with natural language patterns for a fully automated process. Our approach can serve as a baseline for future research on generating UML diagrams from English specifications, and the dataset created together with the defined quantitative metrics can serve as a benchmark for this problem.

The rest of the paper is structured as follows. Section~\ref{Sec:approach} gives an overview of the proposed generation pipeline and the details of each step. The setup and the results of evaluating the approach are provided in Section~\ref{sec:evaluation}. Section~\ref{sec:related} discusses the related work and positions our contribution to it. Section~\ref{sec:validity} lists some threats to validity. Finally, we conclude this paper in Section~\ref{sec:conclusion}.

\section{Approach}
\label{Sec:approach}
\subsection{Overview}

\begin{figure*}[h]
    \centering
    \includegraphics[width=\textwidth]{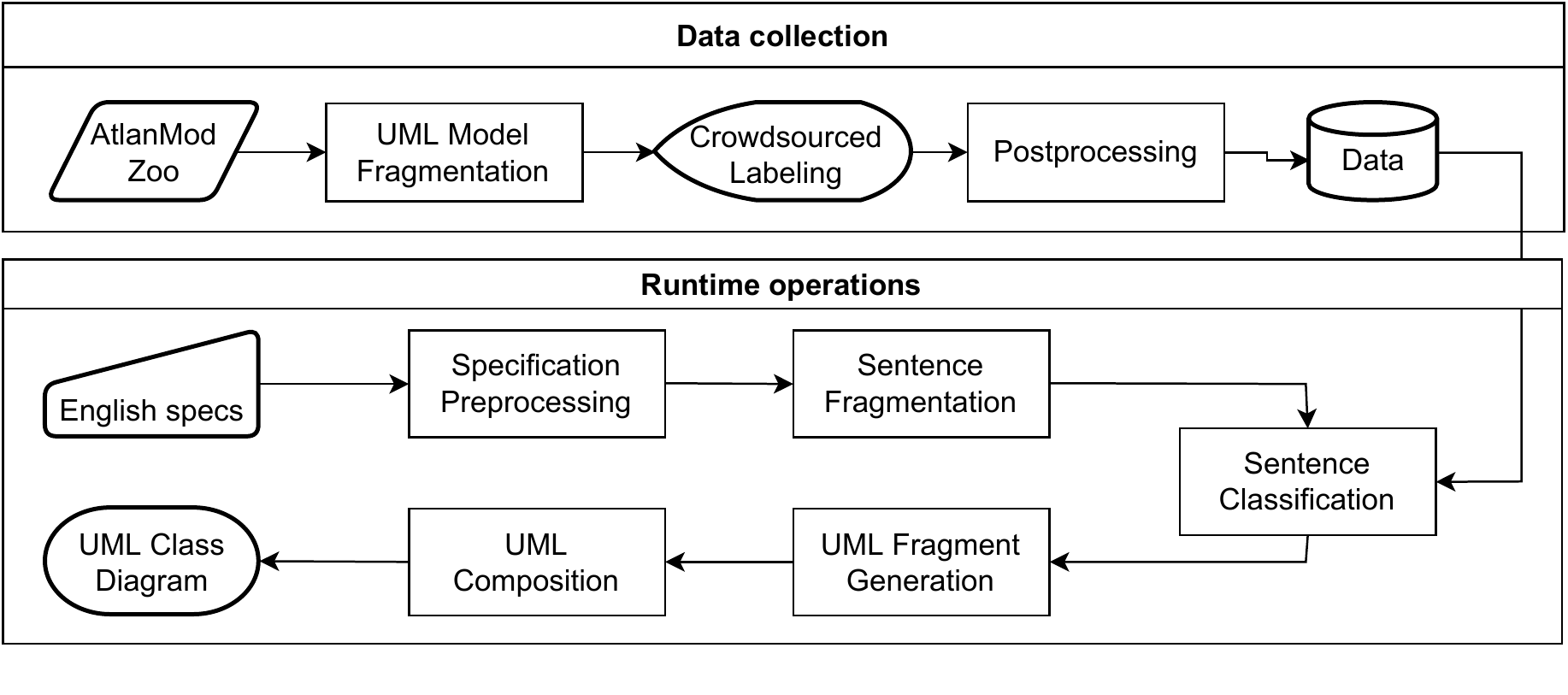}
    \caption{Overview of the extraction process of UML class diagrams from natural language specification}
    \label{fig:overview}
\end{figure*}

The goal is to design a method to translate English specifications to UML diagrams. To do this, we implement a tool pipeline that generates UML class diagrams from natural language specifications. First, we create a dataset. Secondly, we implement an NLP pipeline that performs the extraction of UML class diagrams. Figure 2 summarizes the process. Figure~\ref{fig:overview} summarizes the process.

Our approach combines machine learning with pattern-based diagram generation. To perform machine learning, we start by creating a dataset of UML class diagrams and their corresponding specifications in natural language (top part of Figure~\ref{fig:overview}). We select pre-existing UML class diagrams from the AtlanMod Zoo repository~\footnote{\url{https://web.imt-atlantique.fr/x-info/atlanmod/index.php?title=Zoos}}. The selected diagrams are decomposed into fragments and manually labeled by volunteer participants. After postprocessing, the labeled diagrams are stored in a repository.

The bottom part of Figure~\ref{fig:overview} consists of the actual diagram generation process, which takes place right after a user submits a software specification. The submitted natural language specification is then preprocessed and decomposed into sentences. Using a classifier built from the above-mentioned dataset, the sentences are labeled according to the nature of the UML construct they refer to, i.e. a class or a relation. According to this label, specific procedures of parsing and extraction are performed on the sentence to generate a UML fragment. In the end, all UML fragments are composed back together into one UML class diagram.

\subsection{Dataset Creation} \label{sec:data}
We create a new dataset for both the operation and the evaluation of our approach. In particular, we use this dataset to learn a classifier for the Classification step in Figure~\ref{fig:overview}.

\begin{figure}
    \centering
    \includegraphics[]{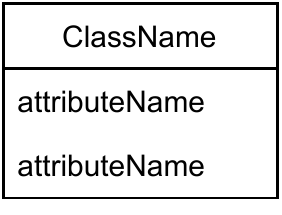}
    \caption{Class fragment}
    \label{fig:class_fragment}
\end{figure}

\begin{figure}
    \centering
    \includegraphics[]{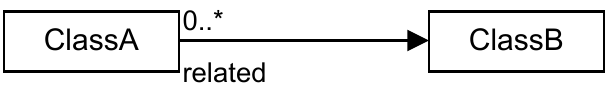}
    \caption{Relationship fragment}
    \label{fig:rel_fragment}
\end{figure}

To build the dataset, we start from an existing set of UML class diagrams from the AtlanMod Zoo. The AtlanMod Zoo has a repository of 305 high-quality UML class diagrams that model various domains. The size of the diagrams varies from a few to hundreds of classes. We fragment each diagram into simple classes (Figure~\ref{fig:class_fragment}) and relationships (Figure~\ref{fig:rel_fragment}). Table~\ref{tab:datasets} shows the size of the initial set of diagrams and the fragments, as well as the portion that we labeled.

\begin{table}[h]
    \centering
    \begin{tabular}{|c|c|c|}
        \hline
        Dataset & UML models & UML fragments \\ \hline
        AtlanMod Zoo & 305 & 8706 \\ \hline
        Labeled & 62 & 649 \\ \hline
    \end{tabular}
    \caption{UML datasets and their sizes by version}
    \label{tab:datasets}
\end{table}

Since we are interested in the translation of specifications into diagrams, each UML class diagram needs to be paired with an English specification. To achieve that goal, we set up a website where we crowdsource the labeling of fragments. The website proposes the labeling of 305 diagrams containing 8706 fragments. We present the diagrams in ascending order of complexity. The website first shows a complete diagram, then iterates on its fragments for labeling while keeping the whole diagram in view. The volunteer participants write an English specification for each fragment. We give examples of labels to help the participants write at the right level of abstraction.

We send the labeling invitation to different MDE mailing lists and specific large research groups active in the MDE field. Volunteer participants are mostly university students and faculty members across the world. To ensure that the labeling is done in good faith, we do not offer monetary compensation for participation. However, since participation was low, we did not impose a contribution limit.

\begin{figure}[h]
    \centering
    
    \begin{subfigure}
        \centering
        \includegraphics[]{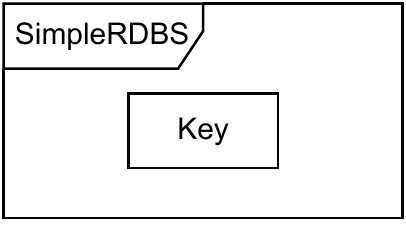}\\
        (a) \textit{Key is a class in SimpleRDBMS package}
    \end{subfigure}
    \newline \\
    \begin{subfigure}
        \centering
        \includegraphics[]{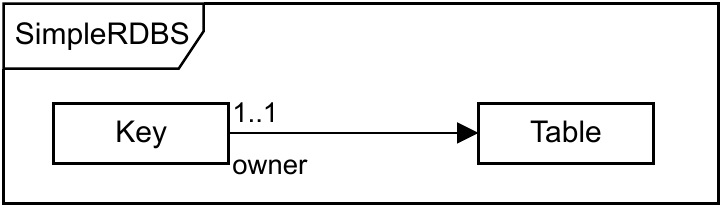}\\
        (b) \textit{A Key is owned by one and only one Table in an RDBMS}
    \end{subfigure}
    \caption{Example labels received by crowdsourcing}
    \label{fig:label_sample}
\end{figure}

After about two months of crowdsourcing, we receive labels for 649 fragments across 62 UML class diagrams. The produced dataset is available on a public repository\footnote{\url{https://github.com/XsongyangX/uml-classes-and-specs}}. 
%In contrast, the website hosted about nine thousand fragments up for labeling. 
To ensure quality, labels are reviewed and some are rejected. We replace the rejected labels by labeling them again ourselves. Figure~\ref{fig:label_sample} shows example labels.

\subsection{Preprocessing and Fragmentation}
Preprocessing is the first step after receiving an input specification from the user as shown in Figure~\ref{fig:overview}. We substitute pronouns throughout the text, such as \textit{it} and \textit{him}, by their reference nouns. This is done using \textit{coreferee}~\cite{coreferee}, which is a tool written in Python that performs coreference resolution, including pronoun substitution.

\begin{quote}
    A course is taught by a teacher. A classroom is assigned to it.\\
    $\Longrightarrow$
    A course is taught by a teacher. A classroom is assigned to a course.
\end{quote}

%\begin{quote}
%    John went home because \textbf{he} was tired.\\
%    $\Longrightarrow$
%    John went home because \textbf{John} was tired.
%\end{quote}

Pronoun substitution allows sentences in the English specification to be less dependent on each other for semantic purposes. The accuracy of \textit{coreferee} for general English text is 81\%.

Sentence fragmentation is the second step in the runtime operations in Figure~\ref{fig:overview}. We split the preprocessed text into individual sentences, using \textit{spaCy}~\cite{spacy}. \textit{spaCy} is an NLP library in Python that can be used for various NLP tasks, such as sentence splitting. \textit{spaCy} splits text into sentences by looking at punctuation and special cases like abbreviations. Its decisions are powered by pre-trained statistical models. We use the small English model, which has a good speed and respectable performance. For instance, in the following example, the first two dots are not considered for splitting the sentences but the third dot is.

% \begin{quote}
%     A class ImplementableStandard. TechnologyObject extends ImplementableStandard on a 1 to many basis.\\
%     $\Longrightarrow$\\
%     $s_1$: A class ImplementableStandard.\\
%     $s_2$: TechnologyObject extends ImplementableStandard on a 1 to many basis.
% \end{quote}

\begin{quote}
    An employee has a level of studies, i.e., a degree. An employee is affiliated to a department.\\
    $\Longrightarrow$\\
    $s_1$: An employee has a level of studies, i.e., a degree.\\
    $s_2$: An employee is affiliated to a department.
\end{quote}

\subsection{Sentence Classification} \label{sec:classification}
Sentence classification is the third step in the runtime operations of Figure~\ref{fig:overview}. Classification provides additional information on the English specification that can be used later to better generate the related UML diagram fragment. Each sentence is classified as describing either a "class" or a "relationship".

The training data for the classifier comes from the dataset described in Section~\ref{sec:data}. Each data point is structured as a pair <English specification, UML fragment> and is assigned a label of a "class" or "relationship" from the moment the dataset was processed from AtlanMod Zoo. The pairing means that the English specification belongs to that specific UML fragment. Our classifier is trained to predict the "class/relationship" label from an English specification. To evaluate the accuracy of the classifier, we use 80\% of the data for the training, and the remaining 20\% for testing.

When training classifiers on natural language text, we have to select a method to map those sentences into numerical representations. To this end, we experiment with two vectorization methods, \textit{count} and \textit{tf-idf}, which are designed to turn words into vectors. \textit{tf-idf}'s key difference from \textit{count} is the penalization of very common words in a document. This allows giving less importance to words like "the".

As for the classification algorithms, we experiment with various algorithms from the \textit{scikit-learn} library~\cite{scikit-learn}. We use the default hyperparameter settings for each algorithm. Table~\ref{tab:classifier-performance} shows the performance of the algorithms on the test data. It is worth noting that the training takes less than one minute. 

\begin{table}[h]
    \centering
    \begin{tabular}{|c|c|c|}
        \hline
         & \multicolumn{2}{|c|}{Vectorizer} \\ \hline
        Classifier & tf-idf (\%) & Count (\%) \\ \hline
    \textbf{Bernoulli Bayes}     &   \textbf{87}  &  83 \\ \hline
    Multinomial Bayes   &   83  &  85 \\ \hline
    k neighbors        &   82  &   74 \\ \hline
    Linear SVC          &   88  &   84 \\ \hline
    SVC                 &   88  &   55 \\ \hline
    ADA                 &   85  &   85 \\ \hline
    Random Forest       &   81  &   70 \\ \hline
    Logistic Regression &   86  &   85-95 \\ \hline
    \end{tabular}
    \caption{Accuracy of binary classification of English sentences}
    \label{tab:classifier-performance}
\end{table}

Although some classifiers have better accuracy, we pick the Bernoulli Naive Bayes classifier with a tf-idf vectorizer. Bernoulli Naive Bayes is simple, has a good accuracy that is more stable across training experiments and is generally faster to execute.

Interestingly, Bernoulli Naive Bayes performs better on a tf-idf vectorizer than the count vectorizer, when it should perform equally well on both in theory. We attribute the difference in performance to the randomized splitting of the dataset when training and testing. Moreover, if a Bernoulli distribution captures enough information to classify well, it seems frequency-based vectorization is not needed.

Should a given sentence may describe both a UML class and a UML relationship, we let the classifier make the decision.

\subsection{UML Fragment Generation} \label{sec:generation}
After classifying each sentence as describing either a "class" or a "relationship", we generate the corresponding UML fragment according to this classification, which is the fourth step in the runtime operations of Figure~\ref{fig:overview}.

Using \textit{spaCy}'s small English model~\cite{spacy}, we define several grammar patterns to match the English sentences. We design the patterns based on the data we collected through crowdsourcing. We broadly group the patterns in Table~\ref{tab:grammars-summary}. For sentences labeled as class descriptions, we define eight patterns $CP1$ to $CP8$, and for those describing relationships, we define six patterns $RP1$ to $RP6$. The patterns make use of part-of-speech tagging and dependency analysis.

\begin{table}
    \centering
    \begin{tabular}{|l|l|}
        \hline
        \textbf{Pattern}    &   \textbf{Example of matching sentence} \\ \hline
        \multicolumn{2}{|c|}{Class fragments}  \\ \hline
        CP1: Copula  &   \textit{Key is a class in SimpleRDBMS package} \\ \hline
        CP2: "there is"  &   \textit{There is a place.} \\ \hline
        CP3: Compound noun  &   \textit{Drawing Interchange Format} \\ \hline
        CP4: Compound explicit   &   \textit{Workflow State class} \\ \hline
        CP5: "to have"   &   \textit{a Mesh has a name of type String} \\ \hline
        CP6: "class named"   &   \textit{A class named "Actor".} \\ \hline
        \multirow{2}{*}{CP7: "of package"}
        &\textit{TextualPathExp is part of the}\\ 
        &\textit{package TextualPathExp} \\ \hline
        CP8: "and" clauses   &   \textit{News have titles and links}\\ \hline
        \multicolumn{2}{|c|}{Relationship fragments} \\ \hline
        RP1: "to have"   &   \textit{A MSProject has at least one task.} \\ \hline
        RP2: Passive voice   &   \textit{A news is published on a specific date} \\ \hline
        RP3: "composed"  &   \textit{A node is composed of a label} \\ \hline
        RP4: Active voice    &   \textit{Eclipse plugins may require other plugins} \\ \hline
        RP5: Noun "with" &   \textit{A table with a caption}\\ \hline
        \multirow{2}{*}{RP6: Copula}
        & \textit{In a Petri Net a Place may be the} \\
        & \textit{destination of a Transition} \\ \hline
    \end{tabular}
    \caption{Summary of patterns}
    \label{tab:grammars-summary}
\end{table}

Multiple patterns can overlap and as such, the \textit{spaCy} parser produces several parse trees for the same sentence. For example, in the category of class fragments, the patterns \textit{CP3: compound noun} and \textit{CP4: compound explicit} are likely to be both applied at the same time. In this case, we set the \textit{CP4: compound explicit} pattern at a higher priority and discard the parse tree from \textit{CP3: compound noun}. 

In general, the priority of patterns in the event of multiple parse trees is based on how specific the pattern is and how much information can be acquired in the parse tree. Hence, for relationship fragments, the patterns for passive voice and active voice are so general that they always yield priority to the other patterns.

After a pattern and its parse tree have been chosen, we generate a UML fragment using a specific template.

\begin{figure}[h]
    \centering
    \textit{News have titles and links} \\
    $\longrightarrow$ \\
    \includegraphics[]{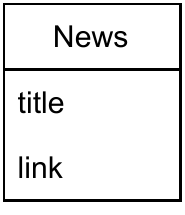}
    \caption{Example of class generation}
    \label{fig:class_generation}
\end{figure}

If the classification of the sentence resulted in "class", we generate a  UML class fragment consisting of only one class with some potential attributes. For example, the \textit{CP8: and clause} pattern creates a class whose name is the subject noun and whose attributes are the objects among the "and" clauses, as shown in Figure~\ref{fig:class_generation}.

\begin{figure}[h]
    \centering
    \textit{A MSProject has at least one task.} \\
    $\longrightarrow$ \\
    \includegraphics{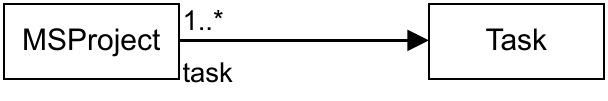}
    \caption{Example of relationship generation}
    \label{fig:relation_generation}
\end{figure}

If the classification of the sentence resulted in a  "relationship", we generate a UML fragment with two classes and one unidirectional association between them. In the case of a "relationship", we can also extract the multiplicity, if present in the sentence. Here is an example with the pattern \textit{RP1: to have} in Figure~\ref{fig:relation_generation}.

During the creation of UML class fragments, we perform additional processing on the parse tree results, such as lemmatization, noun phrase discovery, and variable naming. In Figure~\ref{fig:class_generation}, the attributes are in the singular form of the original word. If Figure~\ref{fig:class_generation}'s input text was "\textit{News have bold titles and url links}", the noun phrase discovery combines the attributes into "boldTitle" and "urlLink" to follow the variable naming convention.

\subsection{Composition of UML Fragments} \label{sec:composition}
The last step in the runtime operations of Figure~\ref{fig:overview} is the composition of UML fragments into one UML class diagram. After each sentence is turned into a UML fragment, we produce the final UML diagram by combining the fragments. Since the merging of general UML class diagrams is NP-hard~\cite{Rubin:2013merging}, we design an algorithm tailored to our use case. The time complexity of our algorithm is polynomial, more specifically $O((c+r)a^2c)$ where $c$ is the number of classes, $r$ is the number of relationships between classes and $a$ is the number of class attributes. 

The composition algorithm takes a greedy approach. Algorithm~\ref{alg:composition} merges one UML fragment at a time into a larger, work-in-progress UML diagram. When all fragments are used, the work-in-process diagram is the completed UML diagram.

\begin{algorithm}
\caption{Composition algorithm}\label{alg:composition}
\begin{algorithmic}[1]
\State $model \gets $ previous composition result or any fragment if the composition is just starting
\State $f \gets $ incoming fragment
\If {kind($f$) = class}
    \If {$\exists c \in model$.classes where $c$.name $= f$.name}
        \If {$\exists$ Attribute-Class conflict}
            \State {resolve Attribute-Class conflict according to Figure~\ref{fig:attribute-class-conflict}}
        \Else {}
        \State {merge attributes from $f$ into $c$ with Algorithm~\ref{alg:merge-attributes}}
        \EndIf
    \Else {}
        \State {insert $f$ into $model$}
    \EndIf
\ElsIf {kind($f$) = relationship}
    \If {$\exists$ Attribute-Relationship conflict}
        \State{resolve Attribute-Relationship conflict according to Figure~\ref{fig:attribute-rel-conflict}}
    \EndIf
    \State \textit{left} $\gets$ class from which $f$ points
    \State \textit{right} $\gets$ class to which $f$ points
    
    % problematic
    \If {\textit{left} $\not\in model$.classes}
        \State{insert \textit{left} into $model$}
    \EndIf
    
    \If {\textit{right} $\not\in model$.classes}
        \State{insert \textit{right} into $model$}
    \EndIf
    
    \If {$f$'s relationship $\in model$.relationships} 
        \State{do nothing}
    \Else {}
        \State{insert $f$ into $model$}
    \EndIf
\EndIf
\State \Return $model$
\end{algorithmic}
\end{algorithm}

\begin{algorithm}
\caption{Merge attributes} \label{alg:merge-attributes}
\begin{algorithmic}[1]
\State {\textit{class} $\gets$ recipient UML class}
\State {$a \gets$ incoming attribute}

\If {$\exists c \in $ \textit{class}.attributes where $c$.name = $a$.name}
    \If {$a$ has a type \textbf{and} $c$ has no type}
        \State {replace $c$ by $a$ inside \textit{class}}
    \EndIf
\Else{}
    \State {insert $c$ into \textit{class}.attributes}
\EndIf
\State \Return \textit{class}
\end{algorithmic}
\end{algorithm}

During composition, fragments may present contradicting information to the model in progress. We identify two situations for this. The first is an Attribute-Class conflict and the second is an Attribute-Relationship conflict (Figure~\ref{fig:attribute-rel-conflict}). In both situations, the resolution involves removing attributes to create a new class or a new relationship. We favor having many smaller classes and relationships, instead of a few very big classes.

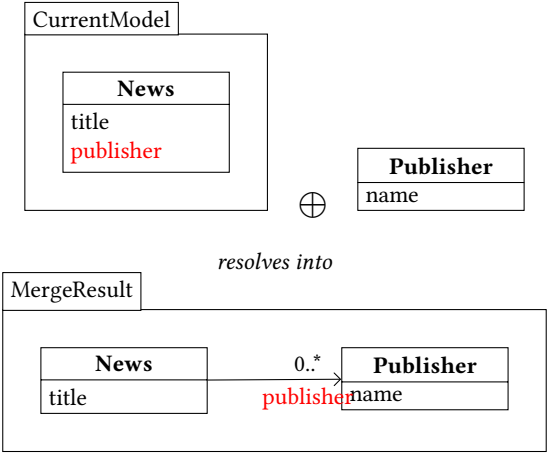
\begin{figure}[h]
    \centering
    \begin{tikzpicture}
        \begin{package}{CurrentModel}
            \begin{class}[text width=2cm]{News}{0,0}
            \attribute{title}
            \attribute{\textcolor{red}{publisher}}
        \end{class}
        \end{package}
    \end{tikzpicture}
    \quad $\bigoplus$ \quad
    \begin{tikzpicture}
        \begin{class}[text width=2cm]{Publisher}{0,0}
            \attribute{name}
        \end{class}
    \end{tikzpicture}
    \\ \quad \\
    \textit{resolves into}\\
    \begin{tikzpicture}
        \begin{package}{MergeResult}
            \begin{class}[text width=2cm]{News}{0,0}
                \attribute{title}
            \end{class}
            \begin{class}[text width=2cm]{Publisher}{4,0}
                \attribute{name}
            \end{class}
        \end{package}
        
        \unidirectionalAssociation{News}{0..*}{\textcolor{red}{publisher}}{Publisher}
    \end{tikzpicture}
    \caption{Attribute-Class conflict and its resolution}
    \label{fig:attribute-class-conflict}
\end{figure}

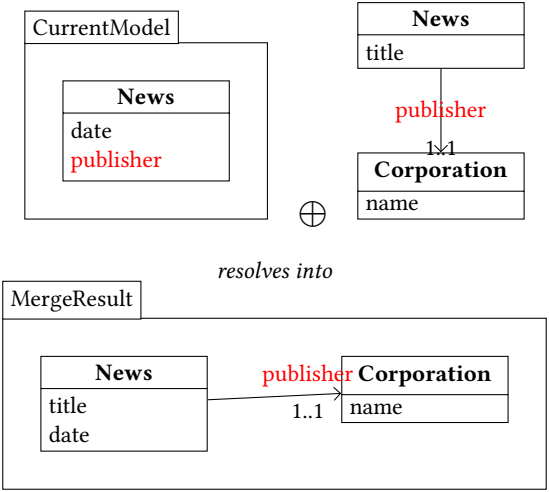
\begin{figure}[h]
    \centering
    \begin{tikzpicture}
        \begin{package}{CurrentModel}
            \begin{class}[text width=2cm]{News}{0,0}
            \attribute{date}
            \attribute{\textcolor{red}{publisher}}
        \end{class}
        \end{package}
    \end{tikzpicture}
    \quad $\bigoplus$ \quad
    \begin{tikzpicture}
        \begin{class}[text width=2cm]{News}{0,0}
                \attribute{title}
            \end{class}
            \begin{class}[text width=2cm]{Corporation}{0, -2}
            \attribute{name}
            \end{class}
        
            \unidirectionalAssociation{News}{\textcolor{red}{publisher}}{1..1}{Corporation}
        
    \end{tikzpicture}
    \\ \quad \\
    \textit{resolves into}\\
    \begin{tikzpicture}
        \begin{package}{MergeResult}
            \begin{class}[text width=2cm]{News}{0,0}
            \attribute{title}
            \attribute{date}
        \end{class}
        \begin{class}[text width=2cm]{Corporation}{4,0}
            \attribute{name}
        \end{class}
        \end{package}
        
        \unidirectionalAssociation{News}{\textcolor{red}{publisher}}{1..1}{Corporation}
    \end{tikzpicture}
    \caption{Attribute-Relationship conflict and its resolution}
    \label{fig:attribute-rel-conflict}
\end{figure}

% formalism of the conflicts
An Attribute-Class conflict arises when the UML class diagram in progress contains an attribute with a name identical to the name of a class from a class fragment. We resolve this conflict by removing the attribute from the diagram in progress, inserting the class fragment into the larger UML diagram, and creating a new relationship from the class that previously contained the attribute to the inserted class. This relationship has the name of the attribute as its name and a multiplicity of zero-to-many. See an example in Figure~\ref{fig:attribute-class-conflict}.

An Attribute-Relationship conflict arises when the UML class diagram in progress contains an attribute $a$ whose name is identical to the name of the relationship inside the incoming relationship fragment. Let $A$ be the class of this attribute $a$ in the diagram in progress. Let $C$ be the class that is the source of the unidirectional association inside the relationship fragment. Let $D$ be the class that is the destination of the unidirectional association inside the relationship fragment. If the names of $A$ and $C$ are not the same, this is not a conflict and we proceed with a standard insertion. Otherwise, the resolution starts by removing the attribute $a$ from $A$. Then we merge the attributes of $A$ and $C$. The relationship from $C$ to $D$ is now from the attribute merge result $A \bigoplus C$ to $D$. 

Lines 16 and 18 in the composition algorithm (Algorithm~\ref{alg:composition}) make use of "relationship" equality. We define two relationships to be equal if the classes they are related to have the same name and if the name of the relationship is the same after processing. This implies that multiplicity is ignored when assessing equality.

Finally, this entire pipeline produces one UML class diagram from the received input. We compile the result into an image using a compiler called \textit{plantuml}~\cite{plantuml}.

\section{Evaluation}
\subsection{Setup}
\label{sec:evaluation}
\begin{figure}[h]
    \centering
    \includegraphics[scale=0.5]{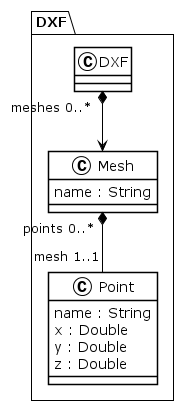}
    \caption{An original UML class diagram from the AtlanMod Zoo}
    \label{fig:original-dxf}
\end{figure}

To test the performance of our approach, we use the dataset we created through crowdsourcing in Section~\ref{sec:data}. We first group all the English specifications for fragments by the UML model they originated from. This creates 62 testing samples. For example, the following grouped specification corresponds to the UML class diagram shown in Figure~\ref{fig:original-dxf}.

\begin{quote}
    \textit{Drawing Interchange Format. a Drawing Interchange model may have multiple meshes. a Mesh has a name of type String. a Mesh may have any number of points. a point maps to only one Mesh. a point has a name of type String and coordinates X and Z of type Double.}
\end{quote}

\subsection{Evaluation Metrics} \label{sec:metrics}
To evaluate the accuracy of our approach, we use comparative metrics. We design three levels of strictness for comparing the diagrams generated by our tool and the ground truth of the dataset. We assume there is only one good UML class diagram per input specification.

First, we have \textbf{exact matching}, which is the most strict comparison. Under exact matching, we look at how many classes and relationships from the ground truth are present in the generated diagrams. A ground truth class is considered present in the prediction if there is a class in the generated diagram that has an identical name and identical attributes. A ground truth relationship is considered "present" in the prediction if both of the classes attached to it are present and if there is a relationship between these two classes with the same name and multiplicity in the predicted output. Each UML diagram under evaluation outputs precision, recall, and f-1 score.

Second, we have \textbf{relaxed matching}, which is a weaker form of exact matching. In relaxed matching, we still look at how many ground truth classes and relationships are present in the generated diagram. However, a ground truth class is considered "present" if there is a predicted class with the same name. We don't look at attributes anymore. Similarly, ground truth relationships are considered "present" in the prediction in the same way as exact matching, except that multiplicities are ignored.

Third, we have \textbf{general matching}. This is the most lenient matching criterion. In general matching, classes are still evaluated like in relaxed matching. Relationships, on the other hand, are evaluated collectively instead of individually. We look at a diagram's graph connectivity and compare this connectivity to the ground truth diagram's connectivity. This comparative metric ignores class names, the orientation of relationships, and the name of relationships. As such, general matching ignores semantics.

To compare two diagrams' connectivity, we use the technique of eigenvector similarity~\cite{eigen_energy}. In short, this technique looks at the eigenvalues of the Laplacian matrices of the two undirected graphs. If the distance between the most prominent eigenvalues is small, the two graphs are similarly connected. Distances are in the range $[0, \infty)$. We apply a mapping to normalize the distance into a score in the interval $(0,1]$, where 0 means no similarity at all and 1 means the two diagrams are identically connected. The mapping is $f(x) = 2(1-\sigma(x))$, where $x \in [0, \infty)$ is the distance and $\sigma(x)$ is the sigmoid function. A perfect connectivity score does not mean the two diagrams are identical.

% \begin{figure}[h]
%     \centering
%     \includegraphics{images/mapping function.png}
%     \caption{Mapping positive reals to the [0,1] interval using $f(x) = 2(1 - \sigma(x))$}
%     \label{fig:mapping_function}
% \end{figure}

To complement connectivity, we add a size difference score to the general matching metric. The size difference is evaluated by computing $||s_1 - s_2||$, where $s_i = normalize([$number of nodes, number of edges$])$ of graph $i$. The norm is the Pythagorean distance, and the graphs are the two undirected graphs used in the eigenvector similarity calculation. The norm ranges from 0 to $\sqrt{2}$, with 0 being the best score. We apply the mapping $g(x) = 1 - x/\sqrt{2}$ to make the score fall in the interval [0,1] with 0 being the worst score and 1 being the best score. This means 0 is attributed to graphs with vastly different sizes, and 1 is attributed to graphs with similar sizes. A score of 1 means the size vectors $s_i$ are oriented closely, not that the graphs have the same sizes. To get a better grasp, we look at the precision and recall results for class generation and the connectivity similarity.

\subsection{Results and Discussion}
After generating 62 candidate UML class diagrams from English specifications, we use an automated script to compare the predictions with the ground truth. Each predicted diagram is compared with its ground truth counterpart in the dataset. We take the average of the 62 results for all metrics and present it in Tables~\ref{tab:performance-classes} and \ref{tab:performance-rels}.

\begin{table}[h]
    \centering
    \begin{tabular}{|c|c|c|c|}
        \hline
        Metric & Precision & Recall & F-1 score \\ \hline
        Exact matching & 0.171 & 0.251 & 0.200 \\ \hline
        Relaxed matching & 0.355 & 0.506 & 0.409 \\ \hline
        General matching & 0.355 & 0.506 & 0.409 \\ \hline
    \end{tabular}
    \caption{Performance of generating classes from English specification}
    \label{tab:performance-classes}
\end{table}

\begin{table}[h]
    \centering
%     \begin{tabular}{|c|c|c|c|}
%         \hline
%         Metric & Precision & Recall & F-1 score \\ \hline
%  %       Exact matching & 0 & 0 & 0 \\ \hline
%         Relaxed matching & 0.140 & 0.098 & 0.110 \\ \hline
%     \end{tabular}\\
%     \quad \\
    \begin{tabular}{|c|c|c|}
        \hline
        Metric & Connectivity similarity & Size difference \\ \hline
        General matching & 0.639 & 0.673 \\ \hline
    \end{tabular}
    
    \caption{Performance of generating relationships from English specification}
    \label{tab:performance-rels}
\end{table}

The evaluation results are not great. The performances are below 50\%, which makes our approach not suited for practical use. We explore several reasons for this and the meaning of low results in the challenge of extracting UML class diagrams from natural language specifications.

% specification synonyms
Since the precision for class generation is low at 17\% for exact matching and 35\% for relaxed matching, it might be caused by too many classes in the prediction that can be traced to synonyms in the specification. English specifications can contain synonyms and different wordings for the same idea. If the user decides to use two different terms for the same concept in a specification, they might have wanted two different classes, or the user might have wanted a specification that is more interesting for humans to read. This ambiguity cannot be resolved without user feedback. However, given our approach generates too many classes on average, a more aggressive merging during the composition step (Section~\ref{sec:composition}) would be beneficial.

% noun phrase detection
While our approach generates too many classes, recall for class generation is still too low at 25\% for exact matching and 50\% for relaxed matching. This means there are elements in the ground truth UML class diagram that are not extracted from English specifications. This can be improved by adding more patterns to the rule set in Section~\ref{sec:generation}. Moreover, a better noun phrase extraction mechanism can extract more class names and attributes from the text. Currently, our method works well with noun phrases that are one word or two words long. Noun phrases longer than two words require a more sophisticated extraction process. We used \textit{spaCy}'s default noun phrase detector, but exploring the dependency parse tree ourselves directly might be a better idea.

% limits of the statistical approach
A low performance signals that our approach has limits. Since we incorporate several statistical components with their own imperfect accuracy in our pipeline, there is an upper ceiling of performance we cannot exceed. Our performance cannot be better than the performances of these components. If we assume all components have an equal influence on the output, we have an expected upper limit of accuracy of 0.63 as seen in Table~\ref{tab:upper-limit}. One way of reducing the effect of compounding errors is to introduce a retroactive step. In our case, that step is the composition algorithm's attempt to resolve conflicts in Algorithm~\ref{alg:composition}.

\begin{table}[h]
    \centering
    \begin{tabular}{|c|c|}
        \hline
        Statistical component & Accuracy \\ \hline
        Pronoun resolution (\textit{coreferee}) & 0.81 \\ \hline
        Grammar analysis (\textit{spaCy}) & 0.90 \\ \hline
        Binary classification & 0.87 \\ \hline \hline
        \textbf{Combined product} & \textbf{0.63} \\ \hline
    \end{tabular}
    \caption{Accuracies of each statistical component in our approach}
    \label{tab:upper-limit}
\end{table}

\section{Threats to Validity} \label{sec:validity}
Although we rejected bad labels during the creation of the dataset, some volunteers provided specifications with questionable semantics and spelling errors. We kept those labels because we want our approach to operate under imperfect conditions. Our approach cannot deal with spelling errors and confusing specifications. Each spelling mistake creates an extra UML class or relationship that should not exist. And if the specification is unclear, then the generated diagram is also unclear.

Due to a lack of sufficient data, we did not set aside unseen data for the evaluation. The evaluation uses the entire dataset for testing. Despite the classifier of Section~\ref{sec:classification} splitting the data into 80-20, 80\% of the evaluation data has been seen during the training of the classifier. We believe this bias to be minimal because the classification is an intermediate step.

In the composition algorithm (Algorithm~\ref{alg:composition}), the merging of the UML class diagram with fragments is treated in a non-commutative fashion. In other words, the pseudo-code only addresses the conflicts when it is an Attribute-Class (Figure~\ref{fig:attribute-class-conflict}) and not Class-Attribute. A similar situation is happening for the Attribute-Relationship conflict. If the conflicting relationship is already inside the model in progress, the algorithm will not flag that as an Attribute-Relationship conflict and it will therefore not resolve it. An improved version of the algorithm should treat the merging of the UML class diagram and the fragment in two directions, i.e. in a commutative way. This would increase the performance of our approach.

\section{Related Work}
\label{sec:related}
\subsection{Survey of Relevant Literature}
In 2021, 24 published tools and methods for the extraction of UML class diagrams from natural language specifications have been surveyed. In the survey, most tools and methods require consistent user intervention. Most tools also required the specification to be given in a specific format, such as a more restricted vocabulary of English or a more rigid structure rather than free-flowing text. The authors concluded that no fully automated tool to generate complex UML class diagrams exists~\cite{Abdelnabi:2021}.

\begin{table}[h]
    \centering
    \begin{tabular}{|c|c|}
        \hline
        \textbf{Degree of automation} & \# of papers \\ \hline
        Semi-automatic & 9 \\ \hline
        Automatic & 15 \\ \hline
    \end{tabular}
    \qquad
    \begin{tabular}{|c|c|}
        \hline
        \textbf{Input} & \# of papers \\ \hline
        Unrestricted English & 12 \\ \hline
        Restricted English & 8 \\ \hline
        Structured format & 4 \\ \hline
    \end{tabular}
    \caption{Results of the 2021 survey on other English-to-UML approaches~\cite{Abdelnabi:2021}}
    \label{tab:uml-survey}
\end{table}

The survey uses qualitative methods to evaluate the outputs of UML extractors. Though valuable, qualitative evaluation is not enough to assess the correctness of the proposed approach and to compare them. We provide quantitative metrics and a testing dataset that can be used for all future UML class diagram extractors.

\subsection{Automatic Approaches}
Automatic approaches do not require extensive user intervention. Once input is given, the user only needs to wait for the recommended result. The automatic approaches make use of more traditional NLP techniques, such as hand-written rules and grammar parsing. For example, the authors of~\cite{Abdelnabi:2020generating, More:2012} use several heuristics to analyze the natural language specification.

The automated approaches presented in \cite{Abdelnabi:2020generating, More:2012} have normalization rules, which require users to write specifications in a restricted English sentence structure. Our approach accepts free-flowing text. We don't have any normalization rules that users must keep in mind.

\subsection{Semi-automatic Approaches}
Semi-automatic approaches make use of an AI assistant to guide the user in general UML class diagrams. DoMoBOT is an example of such a tool. 
%The tool has a complex architecture. 
Overall, DoMoBOT makes use of machine learning via knowledge bases inside pre-trained models and word embeddings. The UML class diagram is generated progressively as the user provides feedback to DoMoBOT~\cite{Saini:2022}.

Although human intervention during the generation of UML models improves quality, the additional effort spent by users makes the tools difficult to use, especially by domain experts. Our approach is meant to improve the automation of the generation process as much as possible. An automated approach is also better for consistent testing.

% \subsection{Model Completion}
% Model generation is one end of a large spectrum. The other end being complete manual modeling. Between the two, model completion offers another way to support domain specialists in their modeling activities. Unlike model generation that use natural language specifications as input, model completion provides targeted recommendations of model elements starting from an existing model fragment. An interesting work in that direction is the one by Elkamel et al. \cite{elkamel_uml_2016} in which the authors use clustering and a similarity metric to recommend model elements. More recently, Burgue\~{n}o et al. \cite{burgueno2021nlp} proposed a general embedding-based approach to suggest concepts in design models. Similarly, Weyssow et al.~\cite{weyssow2022recommending} proposed an approach to suggest concepts in metamodel activities. Their approach consists of retraining an existing deep learning model (RoBERTa) on a limited set of metamodels. In the same vein, Di Rocco et al.~\cite{dirocco2021} used graph neural networks to learn recommenders for model and metamodel completion. Finally, Capuano et al.~\cite{capuano2022} use code repositories to learn a model completion recommender.

% Although the above-mentioned approaches cannot be used directly for model generation, they can be associated with generation tools to improve the quality of the generated diagrams.

\section{Conclusion}
\label{sec:conclusion}

In this paper, we propose an automated approach to extract UML class diagrams from English specifications. The approach uses machine learning and pattern-based techniques. Machine learning is used in the form of a binary classifier that labels sentences as either describing a class or a relationship. The pattern-based techniques are handwritten grammar rules to parse English sentences. In this approach, we fragment the English input into sentences, generate UML class diagram fragments from them, and combine all the fragments together into a final result.

To develop our tool, we first create a dataset of UML diagrams paired with English specifications. The specifications are produced by a crowdsourcing initiative. The resulting dataset, although small, is enough to train the classifier and evaluate our approach. 

We define three evaluation metrics of varying strictness to test our approach's accuracy in generating classes and relationships from an English specification. The results for classes are 17\% precision and 25\% recall for exact matching, the strictest metric. The results for relationships are a connectivity similarity of 63\% and a size difference of 67\%.

The correctness of the produced diagrams is limited. However, these results are in part explained by the imprecision of the NLP tools we used. Using more sophisticated NLP tools will help to improve these results. In addition, more grammar patterns can be added in Section~\ref{sec:generation} and an improved version of the composition algorithm will reduce irrelevant classes.

From a broader perspective, our research lays the work for a consistent quantitative evaluation framework with our approach being the baseline and with the dataset and metrics being the testing framework. From the novelty perspective, we explore intermediate machine learning steps to simplify a mostly rule-based approach. Furthermore, our approach uses a divide-and-conquer strategy when fragmenting diagrams and text and when composing them back together.

In the future, a more complex pattern system can improve the performance of our approach. Currently, we only use a single rule to generate a UML fragment, but if several rules contribute together, the performance can increase. The composition algorithm can also be improved, such as by considering a confidence score in each fragment. Furthermore, inheritance can be generated as a new type of relationship by adding more grammar patterns. Finally, we can generalize our approach to handle other types of UML diagrams, in particular behavioral ones.

\bibliographystyle{ACM-Reference-Format}
\citestyle{acmnumeric}
\bibliography{references}

\end{document}